\documentclass{elsart}
\usepackage{graphicx,amssymb,amsmath}
\usepackage{graphics,graphicx,dcolumn,bm,fleqn,epic,eepic,float}
\usepackage{amssymb,amsmath,multirow,rotate,color,float,times}
\usepackage{color}
\usepackage{citesort}
\definecolor{red}{rgb}{1,0,0}
\definecolor{green}{rgb}{0,1,0}
\definecolor{blue}{rgb}{0,0,1}

\begin{document}
\begin{frontmatter}

\title{Modeling slow deformation of polygonal particles using DEM}

\author[1]{Andr\'es A.~Pe\~na},
\author[1]{Pedro G.~Lind\corauthref{cor}},
\corauth[cor]{Corresponding author}
\ead{lind@icp.uni-stuttgart.de}
\author[3,4]{Hans J.~Herrmann}

\address[1]{Institute for Computational Physics,
            Universit\"at Stuttgart,\\ 
            Pfaffenwaldring 27, 
            D-70569 Stuttgart, Germany,}
\address[3]{Computational Physics, 
            IfB, HIF E12, ETH H\"onggerberg, CH-8093 Z\"urich,
             Switzerland,}
\address[4]{Departamento de F\'{\i}sica, Universidade Federal do Cear\'a,\\
             60451-970 Fortaleza, Cear\'a, Brazil.}

\begin{abstract}
We introduce two improvements in the numerical scheme to
simulate collision and slow shearing of irregular particles.
First, we propose an alternative approach based on
simple relations to compute the frictional contact forces.
The approach improves efficiency and accuracy of the
Discrete Element Method (DEM) when modeling the dynamics 
of the granular packing.
We determine the proper upper limit for the integration step 
in the standard numerical scheme using a wide range of material parameters. 
To this end, we study the 
kinetic energy decay in a stress controlled test between two particles.
Second, we show that the usual way of defining the contact plane
between two polygonal particles is, in general, not unique which
leads to discontinuities in the direction of the contact
plane while particles move.
To solve this drawback, we introduce an accurate definition 
for the contact plane based on the shape of
the overlap area between touching particles, which evolves
continuously in time.
\end{abstract}

\begin{keyword}
Granular media \sep 
Discrete Element Method \sep
Slow deformation \sep
Contact forces.

\PACS 83.10.-y \sep 
      45.70.-n \sep 
      45.70.Cc \sep 
      02.60.-x  
\end{keyword}
\end{frontmatter}


\section{Motivation and model}
\label{sec:model}

To model the dynamics of granular media a commonly used approach is 
the well-known Discrete Element Method (DEM) 
\cite{poeschel05,ciamarra05,dacruz05}.
When applied to slow
shearing~\cite{cundall89,thompson91,pena07} the computation
of frictional and therefore contact forces between particles
may introduce numerical errors that can be larger than 
the precision of the integration scheme used in DEM~\cite{pena07sub}.
Moreover, if irregular particles are considered -- typically 
polygons -- the definition of the contact plane between touching particles 
is not straightforward.
\begin{figure} [t]
\begin{center}
\includegraphics*[width=14.0cm,angle=0]{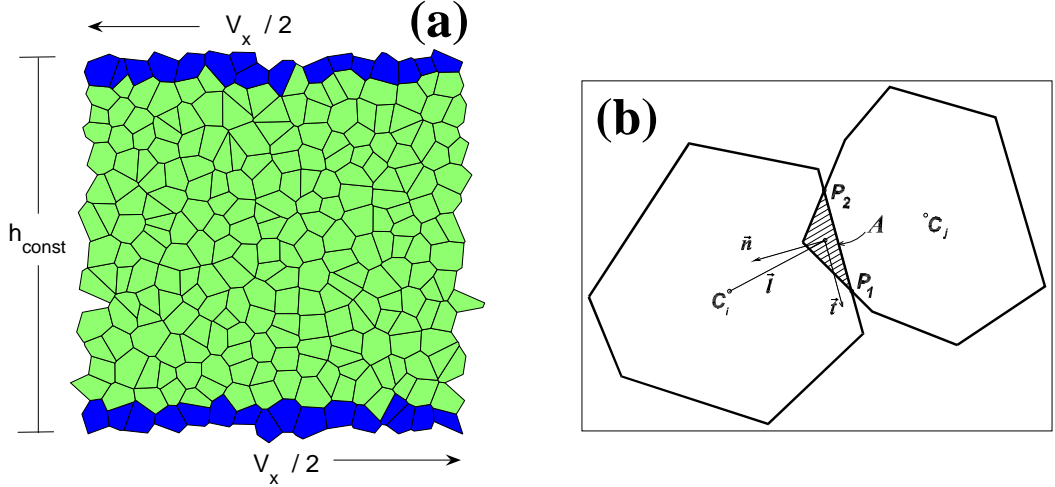}
\end{center}
\caption{\protect
     {\bf (a)} Sketch of the system of $256$ particles (green particles)
     under shearing of top and bottom boundaries (blue particles). 
     Horizontally periodic boundary conditions are considered and 
     a constant low shear rate is chosen.
     {\bf (b)} Illustration of two overlapping particles, where
     the overlap region $A$ between particles fully characterizes the
     contact force $\vec{F}^c$.} 
\label{fig1}
\end{figure}

In this paper we propose an approach to compute the tangential
contact force with the same numerical precision
as the normal force. 
Our improved procedure is specially suited for the case of slow
shearing when large integration steps are needed for efficient
computation, when studying e.g.~the occurrence of 
avalanches~\cite{mora99} and the emergence of 
ratcheting in cyclic loading~\cite{mcnamara07}.
In addition, we 
discuss the drawbacks of the common procedures for computing
the contact plane between touching particles.
We will show that 
for polygonal (anisotropic) particles it varies discontinuously
in time perturbing the proper convergence to the stationary state
in stress controlled tests.
To overcome this shortcoming we introduce a proper definition which
does not depend on the shape of the touching particles and is
suited for both regular and irregular particles.

Our model considers a two-dimensional system, as sketched in
Fig.~\ref{fig1}a.
Each particle has two linear and one rotational degree of freedom, 
being contained between two boundary plates shearing against each 
other~\cite{fernando06,mora99,tillemans95}.
The volumetric strain is suppressed, i.e.~the position of 
the walls is fixed and there is no dilation. 
Periodic boundary conditions are imposed in the horizontal direction.
The evolution of the system is given by the integration of
Newton's equations of motion, where the resulting 
forces and momenta acting on each particle are given by the 
sum of all contact forces and torques applied on that particle, 
respectively. The boundary particles move with a fixed
shear rate.

The integration of Newton's equation of motion is usually 
done with a predictor-corrector scheme~\cite{poeschel05,allen03},
which consists of three main stages, namely prediction, 
evaluation and correction.
In the prediction stage one extracts for each particle 
the predicted position and acceleration of the center of mass. 
This is done by means of Taylor expansions of the
linear and angular positions, yielding the corresponding
velocities and higher-order time derivatives, as functions
of the current values~\cite{allen03,rougier04}.
During the evaluation stage, the predicted coordinate
of each particle is used to determine the contact force 
$\vec{F}^c_{t + \Delta t}$ at time $t + \Delta t$.
Since the method is not exact, there is a difference between
the acceleration $\ddot{\vec{r}}(t + \Delta t) = 
\vec{F}^c_{t + \Delta t} / m$ and the value obtained in the 
prediction stage, namely 
$\Delta \ddot{\vec{r}} = \ddot{\vec{r}}(t + \Delta t) - 
                        \ddot{\vec{r}}^p(t + \Delta t)$.
Finally, the difference $\Delta \ddot{\vec{r}}$ 
is used in the corrector step to correct the predicted 
position and its time derivatives, using proper weights for each time 
derivative~\cite{allen03}.
The weights depend upon the order of the 
algorithm and the differential equation being solved.
These corrected values are used for the next integration
step $t+\Delta t$.

In our simulations we integrate equations of the form 
$\ddot{\vec{r}} = f(\vec{r},\dot{\vec{r}})$, using a fifth order 
predictor-corrector algorithm that has a numerical error proportional 
to $(\Delta t)^6$ for each integration step~\cite{allen03}.

Typically, the particles can neither break nor deform, i.e.~fragmentation 
is neglected. 
The deformation is usually modeled by letting particles 
overlap~\cite{poeschel05,cundall79}, as illustrated in Fig.~\ref{fig1}b. 
The overlap between each pair of particles is considered to fully 
characterize the contact: the normal contact force is assumed to be 
proportional to the overlap area~\cite{tillemans95} and its direction 
perpendicular to the contact plane, which is usually defined by
the intersection points  between the boundaries of the two 
particles~\cite{tillemans95}.

The contact forces, $\vec{F}^c$, are decomposed into their elastic 
and viscous contributions, $\vec{F}^e$ and $\vec{F}^v$ respectively.
The elastic part of the contact force is simply given
by the sum of the normal and the tangential components,
with respect to the contact plane, namely
\begin{equation}
\vec{F}^e = F^e_n \hat{n}^c + F^e_t \hat{t}^c ,
\label{elastContForc}
\end{equation}
where the normal component reads
\begin{equation}
F^e_n = -k_n A/l_c ,
\label{normalelastforc}
\end{equation}
with $k_n$ the normal stiffness, $A$ the overlap area and 
$l_c$ the characteristic length of the contact, which,
for two particles $i$ and $j$, is given by $l_c=r_i+r_j$ 
with $r_i=\sqrt{A_i/2\pi}$ and $A_i$ the area of the
particle $i$, and similarly for particle $j$.
The tangential force is considered to be proportional to the 
elastic elongation $\xi$ of an imaginary spring -- also called
Cundall-Stack spring~\cite{cundall79} -- at the contact, namely
\begin{equation}
F^e_t = - k_t \xi ,
\label{tangelastforc}
\end{equation}
where $k_t$ is the tangential stiffness and
$\xi$ is the elastic elongation updated as 
\begin{equation}
\xi(t+\Delta t) = \xi(t) + \vec{v}_t^c \Delta t,  
\label{elasticelong}
\end{equation}
where $\Delta t$ is the time step of the DEM simulation, 
and $\vec{v}_t^c$ is the tangential component of the relative velocity 
$\vec{v}^c$ at the contact point.

Having both components of the elastic force,
the tangential elastic elongation $\xi$ 
is updated according to the Coulomb limit condition 
$|F_e^t| = \mu F_e^n$,
with $\mu$ the inter-particle friction coefficient.
If the Coulomb condition is reached, particles 
are in the inelastic regime, and sliding is 
enforced by keeping constant the tangential force $F_e^t$.
Here the tangential elongation
$\xi$ takes its maximum value $\pm \mu k_n A/(k_t l_c)$. Otherwise,
if the contact is in the elastic regime ($|F_e^t| < \mu F_e^n$) the 
elongation $\xi$ can increase in time, following Eq.~(\ref{elasticelong}).

The viscous force $\vec{F}^v$ takes into account the dissipation at 
the contact which is important to maintain the numerical stability 
of the method. This force is calculated as~\cite{cundall79}
\begin{equation}
\vec{F}^v = - m_r \nu \vec{v}^c , 
\label{viscousforce}
\end{equation}
where $m_r$ is the reduced mass of the two touching particles
and $\nu$ is the damping coefficient.

The suitable parameters for using this model are the interparticle 
friction $\mu$, the normal stiffness $k_n$, the ratio $k_t/k_n$ between 
tangential and normal stiffnesses and the ratio $\epsilon_t/\epsilon_n$ 
between tangential and normal restitution coefficients.
The restitution coefficients are defined from
the contact stiffness and damping coefficient~\cite{foerster94}
as~\cite{luding98}
\begin{equation}
 \epsilon_n = \exp \ ( -\pi \eta / \omega)
            = \exp{\left ( -\frac{\pi}{\sqrt{4m_rk_n/\nu_n^2-1}}\right )}
\label{eq:Res_coeff}
\end{equation}
for the normal component,
where $\omega = \sqrt{\omega_0^2-\eta^2}$ is the frequency of 
the damped oscillator, with $\omega_0=\sqrt{k_n/m_r}$ the frequency of 
the elastic oscillator, 
$m_r$ the reduced mass, and 
$\eta=\nu_n/(2m_r)$ is the effective viscosity.

We start in Sec.~\ref{sec:sensitivity} by describing the
dependence of the above 
numerical procedure
on the integration step, showing that the computation of the
frictional forces yields a numerical error larger than the
rest of the calculation. 
To overcome this shortcoming we then describe in 
Sec.~\ref{sec:geometrical} a geometrical improvement whose
associated error is, for any studied case, of the same order as the 
error of the predictor-corrector scheme.
In Sec.~\ref{sec:contplane} we address the particular case
of irregular particles, discussing the most common ways of
computing the contact plane between touching particles and 
proposing a more suitable definition.
Discussions and conclusions are given in Sec.~\ref{sec:conclusions}.

\section{Choosing a proper integration step}
\label{sec:sensitivity}

The entire algorithm above relies on a proper choice of the
integration step $\Delta t$, which should 
neither be too large to avoid divergence of the integration
nor too small avoiding unreasonably long computational time.
In this section we explore the numerical error introduced 
by the frictional force in Eq.~(\ref{elasticelong}) and
determine the range of proper integration steps that guarantee 
convergence of the numerical scheme.

A typical criterion to chose the integration step
$\Delta t$ is to take a value such that
$\Delta t < t_c/5$~\cite{luding98,matuttis98},
where $t_c$ is the characteristic duration of
a contact defined by~\cite{thompson91,luding94,luding98}, 
\begin{equation}
t_c = \frac{\pi}{\sqrt{\omega_0^2-\eta^2}} .
\label{tc}
\end{equation}
While in several cases, one uses an integration step
much smaller than the thresholds above~\cite{thompson91},
for low shear rate, very small integration steps imply a high computational 
effort, and so $\Delta t$ should be chosen close to the threshold
$t_c/5$.
As shown below, the upper threshold (fraction of $t_c$)
below which the numerical scheme converges, strongly depends on 
(i) the accuracy of the approach used to calculate frictional forces 
between particles,
(ii) on the corresponding duration of the contact and 
(iii) on the number of degrees of freedom.

In a previous work~\cite{pena07sub} we have shown that considering
values of $\Delta t$ close to the upper limit $t_c/5$ yields
relaxation times depending on $\Delta t$ for the kinetic energy
$E_k(i)=\tfrac{1}{2}\left (
       m_i\dot{\vec{r}}_i^2+I_i\vec{\omega}_i^2
       \right )$ of each particle $i$,
when particles are subject to a non-zero friction.
In particular, in the absence of friction $\Delta E_k$ 
attains a stationary value, while when friction is present it 
changes monotonically.
\begin{figure}[t]
\begin{center}
\includegraphics*[width=8.5cm,angle=0]{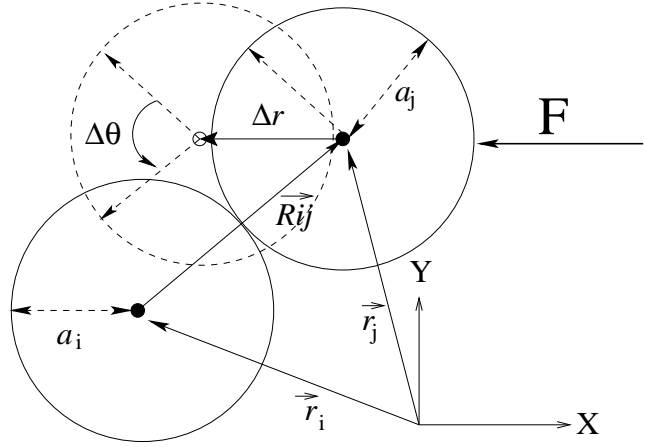}
\end{center}
\caption{\protect
     Sketch of the stress controlled test of two particles (discs).
     The particle located at $R_i$ remains fixed, while the particle
     at $R_j$ is initially touching particle $i$. The vector
     $\vec{R}_{ij}$ connecting the center of mass of particles $i$ and 
     $j$ is initially oriented $45^o$ with respect to the $x$-axis.
     After applying the constant force $\vec{F}$ to disc $j$,
     the system relaxes till it reaches a new position 
     (dashed circumference). 
     Between its initial and final position particle $j$ undergoes 
     a displacement $\Delta R$ and a rotation $\Delta\theta$ 
     (see text).}
\label{fig2}
\end{figure}

To obtain the proper threshold as function of the parameters of our model,
we consider the simple situation of two circular particles in contact,
as sketched in Fig.~\ref{fig2}, and study the kinetic energy of one of 
them under external forcing. 
We start with two touching discs, say $i$ and $j$, where
$i$ remains fixed and $j$ is subject
to a force $\vec{F}$ perpendicular to its surface (no external torque 
is induced) along the $x$-axis.
As a result of this external force, the disc $j$ undergoes translation
and rotation.
The contact forces are obtained from the corresponding springs that are
computed as described in Sec.~\ref{sec:model} and act against the 
external force.
This results in an oscillation of disc $j$ till relaxation (dashed circle 
in Fig.~\ref{fig2}) with a final center of mass displacement of $\Delta R$ 
and a rotation around the center of mass of $\Delta\theta$.

Since $\vec{F}$ is kept constant, the procedure is stress controlled.
Plotting the kinetic energy as a function of time, yields an exponential
decay as
\begin{equation}
E_k(t) = E_k^{(0)}\exp{\left ( -\tfrac{t}{t_R(\Delta t)}\right )} ,
\label{kinen_decay}
\end{equation}
where $t_R$ is a relaxation time whose value clearly depends 
on the integration step $\Delta t$.
Typically~\cite{pena07sub}
the relaxation time $t_R$ depends on $\Delta t$
for integration steps larger than a given threshold $T_t t_c$.
We define this threshold as 
the proper upper limit for the suitable integration steps. 
In other words, we want to determine the upper bound
\begin{equation}
\Delta t \lesssim T_t(\mu,k_t/k_n,\epsilon_t/\epsilon_n) t_c ,
\label{AA}
\end{equation}
where $T_t(\mu,k_t/k_n,\epsilon_t/\epsilon_n)$ is a specific function 
that will be determined below.

\begin{figure}[!t]
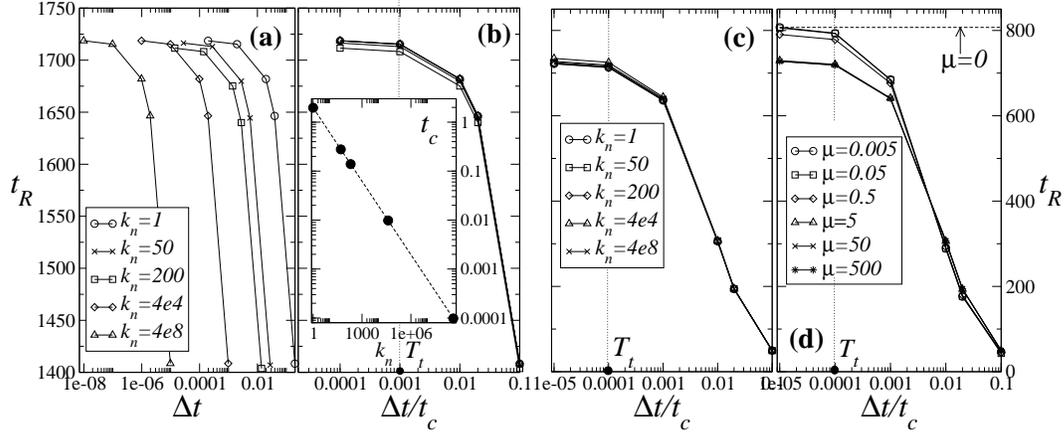

\begin{center}
\includegraphics*[width=6.95cm,angle=0]{fig03a_plh.eps}%
\includegraphics*[width=7.05cm,angle=0]{fig03b_plh.eps}
\end{center}
\caption{\protect
     The relaxation time $t_R$ (in units of $t_c$)
     as a function of
     {\bf (a)} the integration step $\Delta t$ and
     {\bf (b)} the normalized integration step $\Delta t/t_c$,
     where the contact time $t_c$ is defined in Eq.~(\ref{tc}).
     Here the friction coefficient is kept fixed $\mu=500$
     and different stiffnesses $k_n$ (in units of N/m) are considered.
     The quotient $\Delta t/t_c$ collapses all the curves
     for different $k_n$. We find $t_c\sim k_n^{-1/2}$ as 
     illustrated in the inset (see Eq.~(\ref{tc})).
     As a final result one finds a constant $T_t=10^{-3}$.
     For other values of the friction coefficient one
     observes similar results.
     The relaxation time is also plotted as a function of the 
     normalized integration step $\Delta t/t_c$,
     when rotation is suppressed.
     {\bf (c)} $\mu=500$ and different values of $k_n$
     and for
     {\bf (d)} $k_n=4\times 10^8$ and different values of
     $\mu$. The dashed horizontal line in (b) indicates the 
     relaxation time of the kinetic energy in the absence 
     of friction (see text).}
\label{fig3}
\end{figure}

We start by studying the influence of the stiffness, fixing
$\mu=500$ and $\epsilon_t/\epsilon_n=1.0053$.
Figure \ref{fig3}a shows the relaxation time $t_R$ of the kinetic
energy of the two-particle system for $k_n=1, 50, 200, 10^4$ and 
$10^8 \hbox{\ N/m}$.
We see that decreasing $\Delta t$, 
the relaxation time $t_R$ increases till it attains a maximum.
This is due to the computation of the tangential force
using the Cundall-spring scheme in Eq.~(\ref{elasticelong}) 
that yields smaller values for smaller integration steps.
The stabilization of $t_R$ occurs when $\Delta t$ is small
compared to the natural period $1/\omega_0$ of the system.
Thus, $T_t$ is the largest value of $\Delta t$ 
for which we have this maximal relaxation time, in this
case $T_t=10^{-3}$.
Further, all the curves in \ref{fig3}a can be collapsed by 
using the normalized integration step $\Delta t/t_c$,
as shown in Fig.~\ref{fig3}b.
From the inset of Fig.~\ref{fig3}b we also see that the 
contact time scales as $t_c\sim k_n^{-1/2}$, which comes
from Eq.~(\ref{tc}).

\begin{figure}[!b]
\begin{center}
\includegraphics*[width=8.3cm,angle=0]{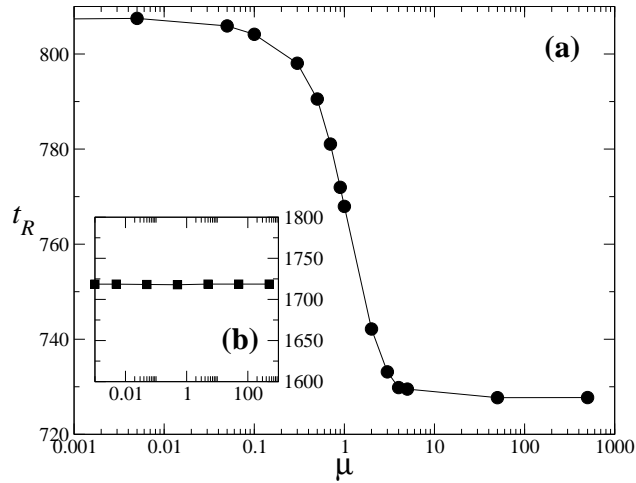}
\vspace{0.5cm}
\end{center}
\caption{\protect
        The relaxation time $t_R$ (in units of $t_c$) as a 
        function of the friction coefficient $\mu$ 
        {\bf (a)} when rotation is suppressed and
        {\bf (b)} when rotation is considered.
        Here, $k_n=4\times 10^8 \hbox{\ N/m}$ which corresponds
        to a contact time $t_c=9.8\times 10^{-5}\hbox{\ s}$. 
        The normalized integration step $\Delta t/t_c=10^{-5}$.}
\label{fig4}
\end{figure}

In the case where rotation is neglected, we obtain a constant 
$T_t=10^{-4}$ as shown in Fig.~\ref{fig3}c.
This $T_t$ value is one order of magnitude smaller than the 
previous one in Eq.~(\ref{AA}) and can be explained by considering 
an `infinite' friction coefficient in the 
rotational degree of freedom~\cite{pena07sub}.
Indeed, by comparing Fig.~\ref{fig3}c with Fig.~\ref{fig3}a, we
see that the relaxation time $t_R$ is smaller when rotation
is suppressed.
However, while Fig.~\ref{fig3}c clearly shows that 
$t_R$ does not depend on the stiffness $k_n$, the same is not true 
for the friction coefficient $\mu$, as shown in Fig.~\ref{fig3}d.

In Fig.~\ref{fig4}a we observe that, in the absence
of rotation, there is a change of the relaxation time around $\mu=1$,
which is not observed when rotation is considered (Fig.~\ref{fig4}b).
This transition occurs since for larger values of $\mu >1$,
one has $F_t > F_n$, and therefore the frictional force increases
and drives the system to relax faster.
Further, with rotation one observes a larger $t_R$ because there is an 
additional degree of freedom, that also relaxes.
\begin{figure}[!b]
\begin{center}
\includegraphics*[width=10.0cm,angle=0]{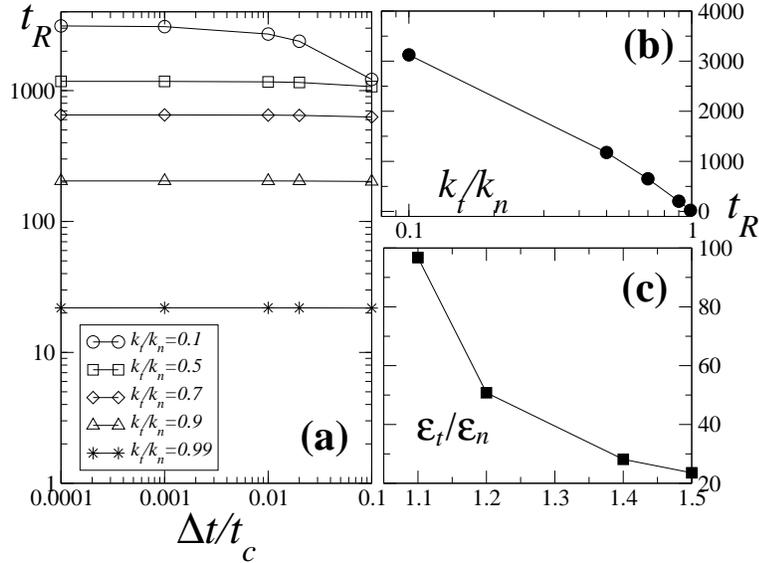}
\end{center}
\caption{\protect 
        Dependence of the relaxation time on the ration
        $k_t/k_n$ of the tangential and normal stiffnesses.
        {\bf (a)} the convergence towards the relaxation time
                  for decreasing integration steps using different
                  stiffness ration.
        {\bf (b)} the approximate logarithmic dependence of $t_R$
                  on $k_t/k_n$ (see text).
        Here $\mu=500$ and $\epsilon_t/\epsilon_n=1.0053$.
        No rotational constraint is here considered.
        Dependence of the relaxation time on the ration
        $\epsilon_t/\epsilon_n$ of the tangential and normal 
        restitution coefficients.
        {\bf (a)} the convergence towards the relaxation time
                  for decreasing integration steps using different
                  restitution ratios.
        {\bf (b)} the relaxation time $t_R$ as a function of
                  $k_t/k_n$ (see text).
        Here $\mu=500$ and $k_t/k_n=1/3$.
        No rotational constraint is here considered.}
\label{fig5}
\end{figure}

We also check the dependence of the relaxation time on the
two other parameters $k_t/k_n$ and $\epsilon_t/\epsilon_n$.
Figure \ref{fig5}a shows the relaxation time as a function
of the normalized integration step $\Delta t/t_c$, using different
stiffness ratios and with fixed $\mu=500$ and $\epsilon_t/\epsilon_n$.
We point out the following.
First the convergence to a stationary value of $t_R$ is faster
for large stiffness ratio as $k_t/k_n\to 1$, yielding larger
values of $T_t$. Second, the stationary value of $t_R$ decreases
with the stiffness ratios. This decrease is logarithmic, as shown
in Fig.~\ref{fig5}b.

The dependence of $t_R$ on the stiffness ratio can be explained
by taking into account that the larger the ratio $k_t/k_n$ the
larger $F_t$.
Since the tangential force $F_t$ controls the 
convergence of the numerical method, 
the larger $k_t/k_n$ the faster the dissipation and thus, the smaller 
the stationary value of $t_R$.
The logarithmic dependence observed in Fig.~\ref{fig5}b 
can be explained from direct inspection of Eq.~(\ref{eq:Res_coeff}).

Figure \ref{fig5}c, showing the relaxation time as a function of 
the ratio $\epsilon_t/\epsilon_n$ for fixed $\mu=500$ and $k_t/k_n=1/3$ 
can be explained similarly as in Fig.~\ref{fig5}b.
Here the dependence of the stationary value of $t_R$ on 
$\epsilon_t/\epsilon_n$ is approximately exponential.

When the Coulomb condition is fulfilled (inelastic regime),
the dependence on $\Delta t$ observed in all figures above, 
does not occur, since the strength of the tangential force 
is given by $F_t=\mu F_n$.
All the previous results are taken within the elastic regime.
This indicates that the improvements in the algorithm
should be implemented when computing the elastic component of the
tangential contact force, in Eq.~(\ref{elasticelong}),
as explained in the next Section.

\section{Improving the integration of the contact force}
\label{sec:geometrical}

Using the Cundall's spring~\cite{cundall89},
the relaxation time of the two particles only converges when 
$\Delta t$ is a small fraction $T_t$ of the contact time $t_c$. 
Such dependence results from Eq.~(\ref{elasticelong}),that includes
$\Delta t$ in the computation of the tangential force in our predictor 
scheme which has an error of $(\Delta t)^2$.

To overcome this shortcoming, we propose a different expression 
to compute the elastic tangential elongation $\xi$. It is based on 
geometric relations and does not use $\Delta t$.
Our expression for $\xi$ contains only the quantities computed 
in the predictor step, guaranteeing a precision for $\xi$ 
of the same order as the one of the predictor-corrector scheme.
We illustrate our approach with the simple system of two discs,
as sketched in Fig.~\ref{fig2}, and discuss further the
more realistic case of polygonal particles.

In Fig.~\ref{fig2}, the elastic elongation of the tangential spring
results from the superposition of both translation and rotational 
degrees of freedom, $\xi_j = \xi_j^{(tr)} + \xi_j^{(rot)}$.

For the translational contribution, the elastic 
elongation $\xi$ depends only on the relative position of the two 
particles.
In this case we substitute Eq.~(\ref{elasticelong}) by the
expression
\begin{equation}
\xi_j^{(tr)}(t+\Delta t) = \xi_j^{(tr)}(t) + \frac{a_i}{a_i+a_j}
     (\vec{R}_{ij}^p(t+\Delta t)-\vec{R}_{ij}^p(t))\cdot \hat{t}^c ,
\label{xi_trans}
\end{equation}
where $a_i$ and $a_j$ are the radii of the discs $i$ and $j$ respectively,
$\vec{R}_{ij}$ is the vector joining both centers of mass
and points in the direction $i\to j$ (see Fig.~\ref{fig2}). 
Index $p$ indicates quantities derived from the coordinates 
computed at the predictor step. 

Considering only the rotational contribution ($\vec{R}_{ij}$ constant),
the elastic elongation $\xi$ depends only on rotation between times 
$t$ and $t+\Delta t$:
\begin{equation}
\xi_j^{(rot)}(t+\Delta t) = \xi_j^{(rot)}(t) + 
                            (\theta_j^p(t+\Delta t)-\theta_j^p(t))a_j ,
\label{xi_rotation}
\end{equation}
where $\theta_j^p(t)$ is the angle of some reference 
point on particle $j$ at the predictor step of time $t$.
\begin{figure}[t]
\begin{center}
\includegraphics*[width=8.3cm,angle=0]{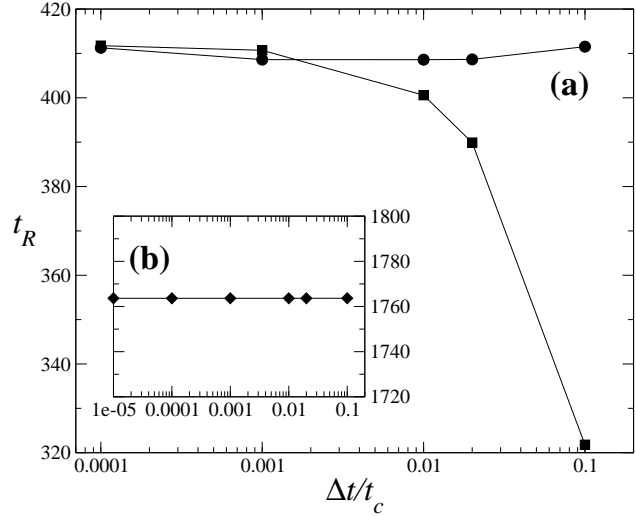}
\end{center}
\caption{\protect 
        {\bf (a)} Comparison of the relaxation time $t_R$ 
        (in units of $t_c$) when using the standard integration scheme 
        (squares) and the proposed improved scheme (circles).
        Here, we perform the stress control test between two particles 
        with irregular polygonal shape, as illustrated in 
        Fig.~\ref{fig1}b and rotation is neglected (see text).
        {\bf (b)} The relaxation time $t_R$ (in units of $t_c$) using 
        Eqs.~(\ref{xi_trans}) and (\ref{xi_rotation}) between
        two discs, as illustrated in Fig.~\ref{fig2};
        for the three cases when considering only rotation, only 
        translation or both, the relaxation time remains constant 
        independent of the integration step (see text).}
\label{fig6}
\end{figure}

Using such expression, we have shown~\cite{pena07sub} that the relaxation 
time $t_R$ is independent on the integration step. This is due to the 
fact that all quantities in the expressions above have an error of the 
same order of the predictor-corrector scheme.

When considering polygonal particles, one must also take into account the 
shape of the particles. 
For polygons, the decomposition of the elastic elongation $\xi$ into 
translational and rotational contributions is not as trivial as 
for discs. This stems from the fact that the contact point no longer 
lies on the vector connecting the centers of mass.
Thus, one should recalculate each time the position of
the center of mass (only from translation) and the relative position
of the vertices (only from rotation).
For that, instead of using Eqs.~(\ref{xi_trans}) and 
(\ref{xi_rotation}), we compute the overlap areas between
the two particles at times $t$ and $t+\Delta t$ and consider the
dislocation of the corresponding geometrical centers.
This yields the translational contribution to
the spring elongation. 
The contribution from the particle rotation is computed by determining 
the angular change due to the rotation of the branch vector between
times $t$ and $t+\Delta t$.

Figure \ref{fig6}a compares how the relaxation time varies with
the normalized time step when the standard Cundall approach is
used (squares) and when our improved approach is introduced
(circles). Clearly, the dependence on the integration step
observed for the usual integration scheme disappears when
our improved approach is introduced.
Therefore, all the conclusions taken above for discs remain 
valid for polygons.
In Fig.~\ref{fig6}b we show the case of two discs for comparison.
\begin{figure}[t]
\begin{center}
\includegraphics*[width=10.0cm,angle=0]{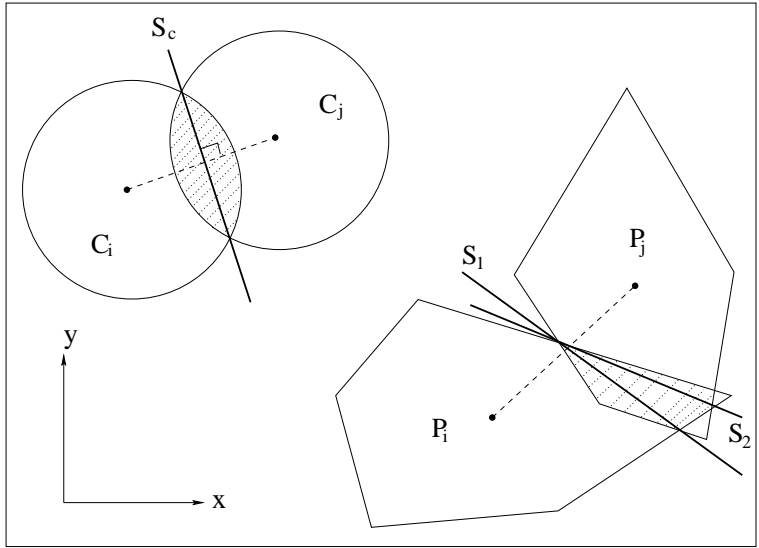}
\end{center}
\caption{\protect 
         Illustration of the definition of contact plane 
         (thick solid lines) between touching discs and
         touching polygons. While for discs the definition is
         unique, since the plane $S_c$ is perpendicular to the
         plain containing the center of the discs, for polygons
         the same is not true, yielding the possibility of more
         than one suitable choice.}
\label{fig7}
\end{figure}

\section{An accurate definition of the contact plane}
\label{sec:contplane}

An additional improvement concerning polygons
is the proper choice of the contact plane.
Usually shear system consider the simple situation of
spherical particles. 
Such particles yield a well defined contact plane, 
namely the vector perpendicular to the one connecting 
the centers of mass of the two touching particles.
Figure \ref{fig7} illustrates this definition of contact plane
for the case of a two-dimensional system.
In the case of polygons the branch vectors are not parallel and therefore
no unique definition is possible.
Usually~\cite{tillemans95} one usually chooses the plane containing 
the intersection points and of the boundaries of the two touching 
particles.
When there are only two intersection points there is a unique
possible contact plane. However, due to the motion of the particles,
the overlap between them may give rise to more than two intersecting
points that yields different possible contact planes.
In Fig.~\ref{fig7} we illustrate a case where two different
choices suit equally the condition above for a contact plane
between two touching polygons.

In this Section, we propose to define the contact plane as the
line passing through the center of mass of the overlap area having
an orientation defined by an angle $\alpha_c$ (with the $x$-axis) 
given by the geometrical average of the orientation 
angles of each (non-oriented) edge of the overlap polygonal with
respect to the $x$-axis.
As shown below such definition is unique and guarantees a continuous 
variation of the orientation of the contact plane while the polygons 
move, as it is the case in a model of irregular particles under
shearing.

More precisely, for an overlap polygon delimited by $m$ edges, where each 
edge $n=1,\dots,m$ lies on a specific line $y=x\tan{\alpha_n}+b_n$, the
contact plane in the two-dimensional case is defined from
\begin{equation}
y = x\tan{\alpha_c} + b_c ,
\label{line}
\end{equation}
where $b_c=y_c-x_c\tan{\alpha_c}$ with $(x_c,y_c)$ the coordinates of  
the center of mass of the overlap polygon, and
\begin{equation}
\alpha_c = \frac{\sum_{n=1}^m \alpha_n \ell_n}{\sum_{n=1}^m \ell_n} ,
\label{averlambda}
\end{equation}
with $\ell_n$ the length of segment $n$.
Since the angles $\alpha_n$ are angles between non-oriented lines,
they can be acute or obtuse. To solve this indetermination we
choose all the angles for the sum in Eq.~(\ref{averlambda}) to be
acute. Then, after computing $\alpha_c$ we choose either the line
in Eq.~(\ref{line}) or the one perpendicular to it, depending 
which one is more perpendicular to the line joining the center
of mass of both particles.
\begin{figure}[!t]
\begin{center}
\includegraphics*[width=10.0cm,angle=0]{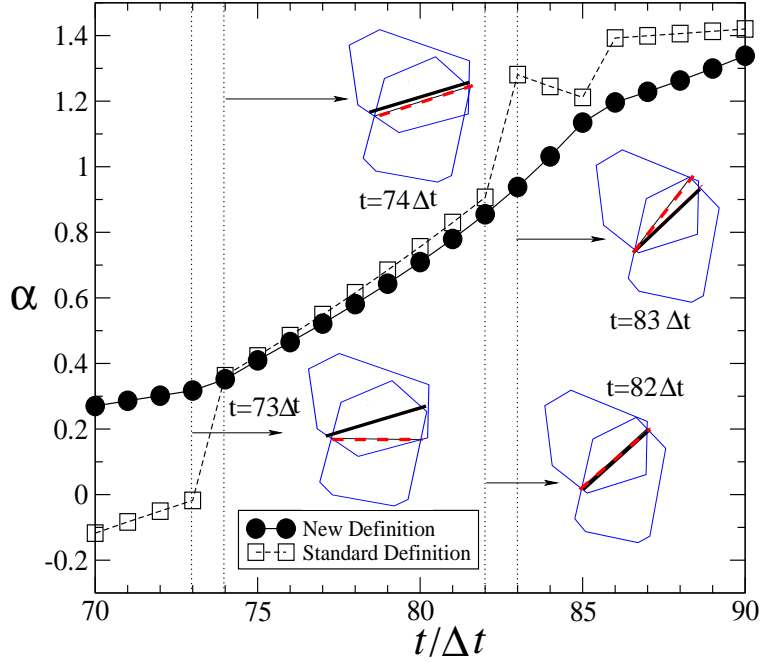}
\end{center}
\caption{\protect 
        Illustration of the orientation of the contact plane,
        given by angle $\alpha_c$ as a function of time when two
        polygons overlap each other.
        Two different definitions of contact plane are
        considered: the usual definition (squares) and the one
        proposed (bullets), as given in Eq.~(\ref{averlambda}).
        One sees that while the proposed definition varies 
        continuously, the usual definition shows discontinuities
        that influence the contact between touching polygons
        (see also Figs.~\ref{fig9} and \ref{fig10}).}
\label{fig8}
\end{figure}
\begin{figure}[!b]
\begin{center}
\includegraphics*[width=13.0cm,angle=0]{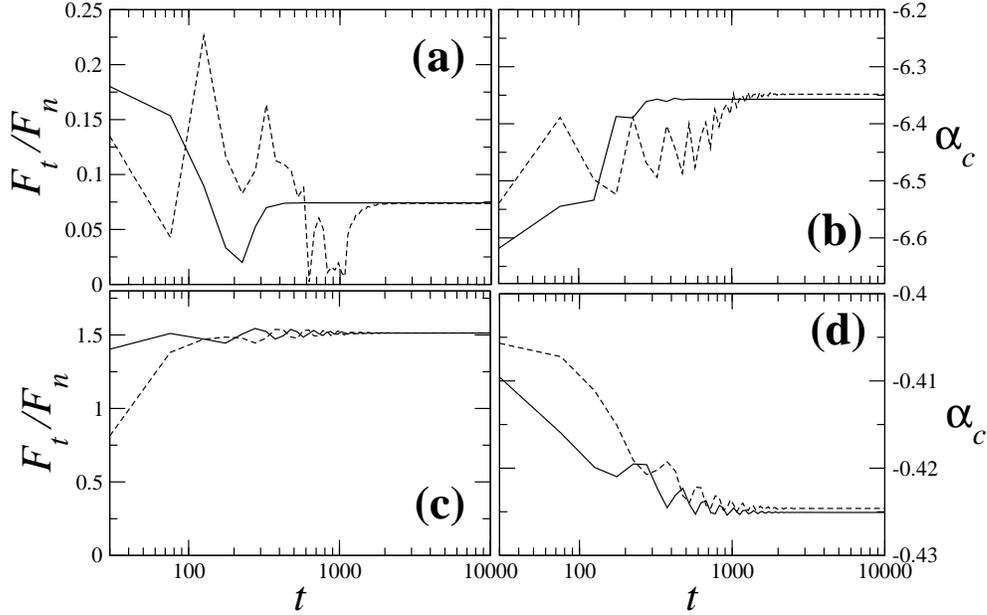}
\end{center}
\caption{\protect 
         Time evolution of 
         {\bf (a)} the ratio $F_t/F_n$ and
         {\bf (b)} of $\alpha_c$ for the usual definition
         of the contact plane.
         The same evolution is plotted for
         {\bf (c)-(d)} the new definition in Eq.~(\ref{averlambda}).
         In each case two curves are plotted, corresponding to two
         different integration steps (see text).}
\label{fig9}
\end{figure}
\begin{figure}[htb]
\begin{center}
\includegraphics*[width=10.0cm,angle=0]{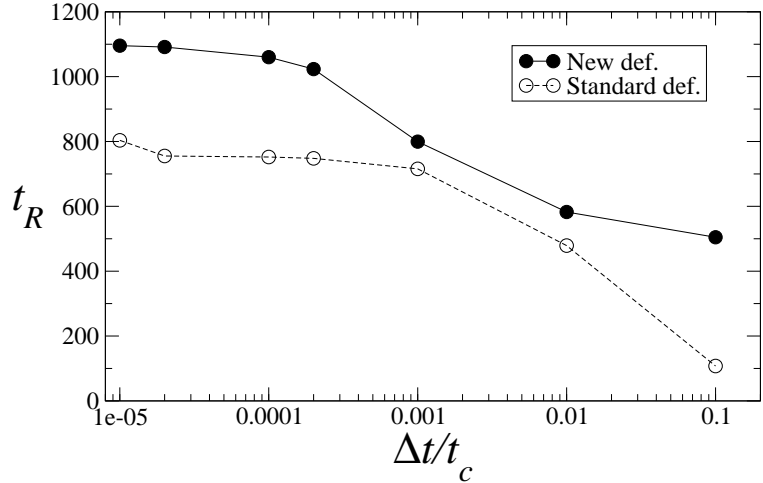}
\end{center}
\caption{\protect 
         The relaxation time as a function of the integration
         step for the usual definition (circles) and the new
         definition of the contact plane (bullets).}
\label{fig10}
\end{figure}

The main advantage of our definition in Eq.~(\ref{averlambda}) is
that while particles move the orientation of the contact plane 
varies continuously with their overlap (polygonal) area, as illustrated
in Fig.~\ref{fig8} with the curve marked by bullets.
On the contrary, contact planes taken from two intersection points of
the two polygons present frequent discontinuities, as can be seen from 
the examples with squares in Fig.~\ref{fig8}. 
Such discontinuities occur since not only the size and orientation
of the edges defining the overlap polygon vary in time, but also the
{\it number} of edges.

There is also the advantage that such a definition enables the system to 
behave as physically expected. For instance, during relaxation of a 
stress controlled test as illustrated in Fig.~\ref{fig9}, the characteristic
physical properties oscillate and converge to a stationary value.
In Fig.~\ref{fig9} we plot both $F_t/F_n$ and $\alpha_c$
for both definitions.
We can see that the convergence of
the new definition is similar to a damped oscillation, contrary to 
what is observed for the conventional definition.
Such deviation from the damped oscillation behavior occurs
due to the discontinuities observed in Fig.~\ref{fig8}, which 
act as external excitations in the damping of the particles
oscillation.
As a consequence,
the relaxation time associated with the discontinuous
contact plane is smaller than when using our definition,
as illustrated in Fig.~\ref{fig10}.

\section{Discussion and conclusions}
\label{sec:conclusions}

We introduced a technique to improve the accuracy of the numerical 
scheme used to compute the evolution of particle systems, showing
that the range of admissible integration steps has an upper limit 
significantly smaller than previously assumed.
The new approach for computing the frictional forces is an alternative
to the Cundall spring, given by Eqs.~(\ref{xi_trans}) 
and (\ref{xi_rotation}) and suits not only the simple 
situation of discs but also more realistic ones, where
particles have polygonal shape.

Further, our upper limit of the admissible range of integration 
steps was determined for a single contact using a stress controlled test. 
Its dependence on the stiffness, restitution coefficient and
friction coefficient was carefully addressed.
When a larger system ($N$ particles) is studied, multiple contacts 
must be taken into account, which will yield an upper limit that we 
conjecture to be $1/N$ of the upper limit obtained above.

Finally, for the case of polygonal particles, we also introduced
a definition for the contact plane between two touching particles.
Our definition is not only based on the intersections points of 
the particles in contact but also on the geometry of the overlap 
area. Thus, the contact plane varies continuously
during the stress controlled tests and is unique, contrary to the 
usual definition.

\section*{Acknowledgments}
The authors thank Sean McNamara and 
Fernando Alonso-Marroqu\'{\i}n for useful discussions.
We thank support by German-Israeli Foundation and
by {\it Deutsche Forschungsgemeinschaft}, under the project
HE 2732781.
PGL thanks support by {\it Deutsche Forschungsgemeinschaft}, under
the project LI 1599/1-1. HJH thanks the Max Planck prize.


\end{document}